\documentclass[prb,showpacs,aps,twocolumn]{revtex4}
\usepackage{amsmath}
\usepackage{graphicx}
\usepackage{dcolumn}
\usepackage{amssymb}

\begin{document}

\title{ Many-body effects in the cyclotron resonance of a magnetic dot}
\author{Nga T. T. Nguyen$^1$}
\email{nga.nguyen@ua.ac.be}
\author{F. M. Peeters$^{1,2}$}
\email{francois.peeters@ua.ac.be} \affiliation{$^1$Departement
Fysica, Universiteit Antwerpen, Groenenborgerlaan 171, B-2020
Antwerpen, Belgium \\$^2$Departamento de F\'{i}sica, Universidade
Federal do Cear\'{a}, Caixa Postal 6030, Campus do Pici, 60455-760
Fortaleza, CE, Brazil}
\begin{abstract}

Intraband cyclotron resonance (CR) transitions of a two-electron
quantum dot containing a single magnetic ion is investigated for
different Coulomb interaction strengths and different positions of
the magnetic ion. In contrast to the usual parabolic quantum dots
where CR is independent of the number of electrons, we found here
that due to the presence of the magnetic ion Kohn's theorem no
longer holds and CR is different for systems with different number
of electrons and different effective electron-electron Coulomb
interaction strength. Many-body effects result in \emph{shifts in
the transition energies} and \emph{change the number of CR lines}.
The position of the magnetic ion inside the quantum dot affects the
structure of the CR spectrum by changing the position and the number
of crossings and anti-crossings in the transition energies and
oscillator strengths.

\end{abstract}

\pacs{78.67.Hc, 71.55.Eq, 75.75.+a, 75.50.Pp}

\maketitle

\section{Introduction.}
Single-doped magnetic impurity quantum dots (QDs)\cite{Furdyna} are
considered as promising nano-structures for spintronic physics.
Besides, these systems allow to probe the impurity spin
states\cite{Besombes,Leger}. Furthermore, spin information in these
spin-based devices can be stored\cite{Gall} (and retrieved) in the
magnetic moment of the Mn-ion. In these nano-systems one can confine
a small number of electrons that can be manipulated by e.g. the gate
voltage. These diluted magnetic semiconductor nano-structures have
attracted a lot of attention to theoreticians and experimentalists
within the past ten years. Those
studies\cite{Besombes,Leger,Govorov,Schmidt,Worjnar,Fernandez,Chang,Gall,Glazov,Hawrylak,NTTNguyen2}
include investigations of the electronic structure, the magnetic
properties, optical absorption, etc. For a recent review of
experimental and theoretical work on QDs doped with magnetic
impurities the reader is referred to Ref.\cite{Govorov_review}

Magnetic-optical properties of nonmagnetic few-electron QDs were
investigated in e.g.
Refs.\cite{Kohn-Luttinger,Kohn,Peeters,Sikorski,Geerinckx,Helle}. It
was found theoretically\cite{Kohn,Peeters} and
experimentally\cite{Sikorski} that the optical cyclotron resonance
energies in the case of quadratic confining potentials are
independent of the number of electrons. In other words, the
electron-electron (e-e) interaction does not result in any
observable changes in the far-infrared (FIR) spectrum for different
$N_e$-electron systems. This theorem is no longer valid in the
presence of impurities as found experimentally\cite{Nicholas} and
theoretically\cite{Merkt}.

In this paper we study a magnetic
semiconductor system [Cd(Mn)Te] in the presence of a Mn-ion of spin
$5/2$ ($Mn^{2+}$). We find that the excitation energy spectrum of
the quantum dot will change due to the presence of the Mn-ion.
Different CR lines in the electron absorption energy spectrum are
found as a consequence of the electron-magnetic-ion (e-Mn) spin-spin
exchange interaction. Positioning the Mn-ion at different positions
inside the QD affects the results significantly. We found
also that the e-e interaction influences the electron excitation
spectrum.

A two-dimensional few-particle circular parabolic QD was considered
as a model system to study e.g. the pair Coulomb interaction
effect\cite{Maksym,Wojs}. Here we will use this model to examine the
effect of the e-e interaction on the intra-band absorption spectrum
of a two-electron QD containing a single Mn-ion. Detailed studies of
the FIR spectra in the case without a Mn-ion were done for quantum
dot molecules\cite{Helle} containing few electrons in which the e-e
interaction leads to small shifts in the peak position of the CR
spectrum.

The intra-band energy absorption spectrum of a magnetic QD was
calculated by Savi\'{c} \emph{et al}\cite{Savic} for a single
electron in the presence of one and two Mn-ions in a 3D
non-parabolic CdTe/ZnTe QD system, and recently by
ourselves\cite{NTTNguyen1} for a single electron in a 2D purely
parabolic single-doped-Mn$^{2+}$ CdTe self-assembled QD. Both
studies pointed out that with the presence of the Mn-ion, the
in-plane intra-band absorption energy spectra exhibits several
CR-lines as a consequence of the e-Mn interaction. Here we extend
our previous work to the case of two electrons and study the effect
of the e-e interaction on the CR and the magneto-optical absorption
spectrum. One way to see that the e-e interaction will affect the CR
spectrum is by writing the two-electron Hamiltonian in terms of
center-of-mass
($\overrightarrow{R_c}=\frac{\overrightarrow{r_1}+\overrightarrow{r_2}}{2},
\overrightarrow{P}=\overrightarrow{p_1}+\overrightarrow{p_2}$) and
relative
($\overrightarrow{r}=\overrightarrow{r_1}-\overrightarrow{r_2},
\overrightarrow{p}=\overrightarrow{p_1}-\overrightarrow{p_2}$)
coordinates:
\begin{equation}\label{e:CM+rel}
H = H_{R_c}+H_{r}+H_{sM}+H_{Z}
\end{equation}
where
\begin{subequations}
\label{eq:whole}
\begin{equation}
H_{R_c} = \frac{1}{2M^{*}}
(\overrightarrow{P}+Q\overrightarrow{A_c})^2 +
\frac{1}{2}M^*\omega_0^2 R_c^2,\label{subeq:1}
\end{equation}
\begin{equation}
H_r=\frac{1}{2M^{*}}
(\overrightarrow{p}+\frac{1}{2}Q\overrightarrow{A_r})^2 +
\frac{1}{2}M^*\omega_0^2{(r/2)}^2+\frac{KQ^2}{4\epsilon
r},\label{subeq:2}
\end{equation}
and
\begin{equation}
H_{sM}=J_c\overrightarrow{M}\cdot[\overrightarrow{s_1}\,\delta(\overrightarrow{R_c}+\frac{\overrightarrow{r}}{2}-
\overrightarrow{R})
+\overrightarrow{s_2}\,\delta(\overrightarrow{R_c}-\frac{\overrightarrow{r}}{2}-\overrightarrow{R})]\label{subeq:3}
\end{equation}
\end{subequations}
are, respectively, the center-of-mass, the relative, and the
electron spin Mn-ion spin Hamiltonian. $H_Z$ is the total Zeeman
spin energy. $M^*=2m^*$ and $Q=2e$ are, respectively, the total mass
and total charge of the electrons. $\omega_0$ is the confining
frequency,
$\overrightarrow{A_c}=\frac{1}{2}(\overrightarrow{B}\times\overrightarrow{R_c})$
and
$\overrightarrow{A_r}=\frac{1}{2}(\overrightarrow{B}\times\overrightarrow{r})$
are the center-of-mass and relative magnetic vector potential,
respectively. In a CR experiment the long-wavelength radiation
interacts only with the center of mass through
$e\overrightarrow{E}\cdot\overrightarrow{R_c}$ where
$\overrightarrow{E}$ is the electric field of the FIR. Notice that
due to the interaction of the electron spin with the Mn-ion spin
$\overrightarrow{M}$ the center of mass is coupled with the relative
coordinates which contain information on the e-e interaction.

In the next section we describe our theoretical approach to solve
the many-particle system which is based on the configuration
interaction (CI). Sect. III presents our numerical results. A
summary and our conclusions are given in Sec. IV.

\section{Theoretical approach}
The Hamiltonian\cite{Hawrylak,NTTNguyen2} for a parabolic quantum
dot containing two electrons interacting with a single Mn-ion
($Mn^{2+}$) in the presence of a perpendicular magnetic field in
second-quantized form reads:
\begin{align}\label{e:secondquantized}
\hat H =& \sum_{i,\sigma}E_{i,\sigma}c_{i,\sigma}^+c_{i,\sigma}
+\frac{1}{2}\hbar\omega_c\left(g_e \frac{m^*}{m_0}
 {S}_z+g_{Mn} \frac{m^*}{m_0}{M}_z\right)
 \nonumber\\
&+\frac{1}{2}\sum_{ijkl\sigma \sigma^{'}} \langle i,j|V_0|k,l\rangle
c_{i,\sigma}^+c_{j,\sigma^{'}}^+c_{k,\sigma^{'}}c_{l,\sigma}
\nonumber\\
 &-\sum_{ij}\frac{1}{2}J_{ij}\left(\overrightarrow{R}\right)
 [ \left(c_{i,\uparrow}^+
 c_{j,\uparrow}-c_{i,\downarrow}^+c_{j,\downarrow}\right)M_z
  \nonumber\\
   &+c_{i,\uparrow}^+c_{j,\downarrow}M^{-} + c_{i,\downarrow}^{+}
c_{j,\uparrow} M^{+} ],
\end{align}
where $i$ in the sum is the single-particle Fock-Darwin state:
$\varphi_{i=\{n_rl\}}\left(\overrightarrow{r}\right)=
\frac{1}{l_H}\sqrt{\frac{n_r!}{\pi\left(n_r+|l|\right)!}}\left(\frac{r}{l_H}\right)^{|l|}
e^{-il\theta}e^{-\frac{r^2}{2l_H^2}}L_{n_r}^{|l|}\left(\frac{r^2}{l_H^2}\right)$
having corresponding on-site energy: $E_{i,\sigma}=\hbar\omega_H
(2n_r+|l|+1)-\hbar\omega_cl/2$ with the frequency $\omega_H=\sqrt
{\omega_0^2+\omega_c^2/4}$ that defines a new length
$l_H=\sqrt{\hbar/m^* \omega_H}$ where $\omega_c=e B/m^*$ is the
cyclotron frequency. This new effective length is related to the
confinement length $l_0$ via the relation:
$l_H=l_0/(1+\Omega^2_c/4)^{1/4}$ with $\Omega_c=\omega_c/\omega_0$.
It is convenient to use this dimensionless parameter $\Omega_c$
instead of the frequency $\omega_c$. For our numerical work we
consider the host semiconductor CdTe and indicate the Land\'{e}
g-factor of the electron and the Mn-ion by $g_e$ and $g_{Mn}$,
respectively. The strength of the electron-Mn-ion spin-spin exchange
interaction is evaluated via:
$J_{ij}(\overrightarrow{R})=J_c\varphi_i^*(\overrightarrow{R})\varphi_j(\overrightarrow{R})$
as the product of single-electron Fock-Darwin wave functions $i$ and
$j$ at the position of the Mn-ion. We define a dimensionless
parameter: $\lambda_C=l_0/a_B$ as the Coulomb interaction
strength\cite{NTTNguyen2} with
$a_B^{*}=4\pi\epsilon_0\epsilon\hbar^2/m^*e^2$ the effective Bohr
radius.

We use the many-body wave function, built up from the Fock-Darwin
basis, namely the Slater determinant for a specific configuration of
the electrons and the Mn-ion, which now includes the spin part of
the Mn-ion $\chi_{\varsigma}\left(\overrightarrow{M}\right)$. As an
example, the two-electron wave function for configuration
$k=\{m,n,\varsigma_{\overrightarrow{M}}\}=\{(n_r,l,s_z)_{m},(n_r,l,s_z)_{n},\varsigma_{\overrightarrow{M}}\}$
is written:
\begin{align}\label{e:WF}
\psi_{k}\left( \overrightarrow{x^*_1},\overrightarrow{x^*_2},
\overrightarrow{M}\right)=\frac{1}{\sqrt{2}}[
\phi_{m}(\overrightarrow{x^*_1})\phi_{n}(\overrightarrow{x^*_2})-
\phi_{n}(\overrightarrow{x^*_1})\phi_{m}(\overrightarrow{x^*_2})]
\chi_{\varsigma}(\overrightarrow{M}),
\end{align}
where
$\overrightarrow{x^*_i}=(\overrightarrow{r_i},\overrightarrow{s_i})$
is the radial and spin coordinates of the $i^{th}$ electron in state
$\phi_{\alpha}\left(\overrightarrow{x^*_i}\right)=
\varphi_{\alpha}\chi_{\sigma}\left(\overrightarrow{s_i}\right)$ - a
product of the Fock-Darwin and the spin coordinate - as the
single-particle wave function. Now we employ the CI method to define
the wave function of the system: $\Psi \left(\overrightarrow{x^*_1},
\overrightarrow{x^*_2}, \overrightarrow{M}\right) = \sum_{k=1}^{N_c}
c_k \psi_k\left(\overrightarrow{x^*_1}, \overrightarrow{x^*_2},
\overrightarrow{M}\right)$ as a linear combination of all possible
configurations $N_c$ where $\psi_k$ is given by Eq. (\ref{e:WF}).
The number of configurations $N_c$ is determined by the number of
Fock-Darwin orbitals $N_s$, the number of electrons which in the
present study is $2$, with taking into account a factor of $2$ due
to the electron spin
$1/2$ and a factor of $6$ due to the Mn-ion spin size $5/2$: $N_c=6\times \begin{pmatrix}2 \\
2N_s\end{pmatrix}$. For our numerical work we use the set of
parameters used in Refs.\cite{Hawrylak,NTTNguyen2} with $m^*=0.106
m_0$, $g_e=-1.67$, $g_{Mn}=2.02$, $a_B^{*}=52.9$ ${\AA}$,
$J_c=1.5\times 10^{3} meV {\AA}^{2}$, and dielectric constant
$\epsilon=10.6$. The effective Coulomb interaction strength
$\lambda_C$ is changed by changing $l_0$, i.e. which is a measure
for the size of the QD, which is typically about tens of angstroms.
$N_c$ is taken sufficiently large to guarantee convergency.

\section{Numerical results}
\subsection{Cyclotron resonance}
The oscillator
strength for circular polarized light is given by:
\begin{equation}\label{e:OS}
 f_{ij}=\frac{2\Delta
 E_{ij}}{\hbar \omega_H}\cdot\frac{|A_{ij}|^2}{l_H^2},
\end{equation}
with the transition amplitude for our two-electron system:
\begin{equation}\label{e:Aij}
 A_{ij}=\sum_{p=1}^{N_e=2}<\Psi_i(\overrightarrow{r_1},\overrightarrow{r_2})
 |r_p e^{\pm i\theta_p}|\Psi_j(\overrightarrow{r_1},\overrightarrow{r_2})>.
\end{equation}
\begin{figure}[btp]
\begin{center}
\vspace*{-0.5cm}
\includegraphics*[width=9.0cm]{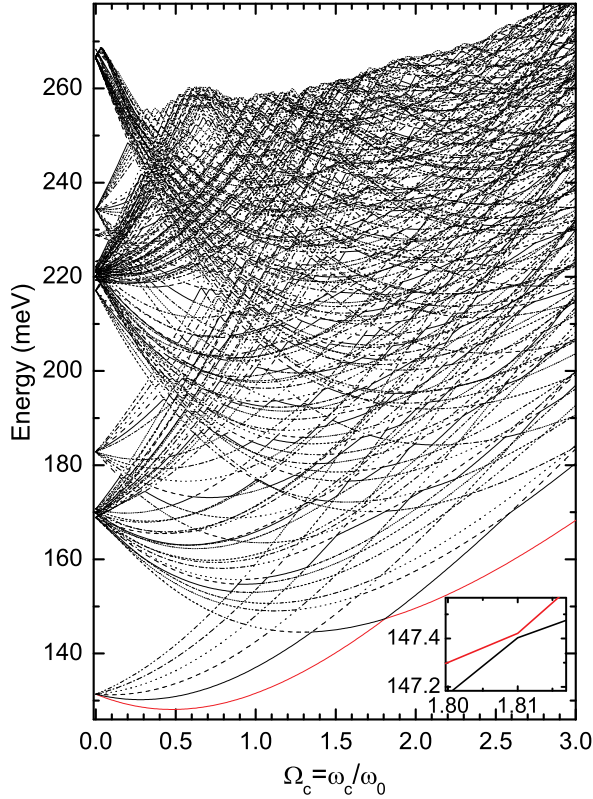}
\end{center}
\vspace{-0.5cm} \caption{(Color online) First $200$ energy levels of
the spectrum of a two-electron magnetic QD with the Mn-ion located
at the center of the QD and for a Coulomb interaction strength
$\lambda_C=0.5$. The inset is the anti-crossing point of the first
two energy levels at $\Omega_c=1.81$}\label{spectrum_cen_lamb1_Ne=2}
\end{figure}

The $\pm$ sign refers to right and left circular polarized light.
$\Delta E_{ij}=E_j-E_i$ is the cyclotron transition energy for the
system from state $i$ to state $j$. Clearly, the spins of the
electrons and the Mn-ion are not involved in the transition. From
expression (\ref{e:Aij}), we see that one of the electrons can
transit to a higher (lower) energy level while the other electron
can be involved in another transition. Therefore, the wave function
part of the other electron can be separated from each term of the
Slater determinant while one electron resonates.

We express the Slater determinant (\ref{e:WF}) in the language of
permutations for identical particles: $\psi_{k} \left(
\overrightarrow{x^*_1}, \overrightarrow{x^*_2},
\overrightarrow{M}\right)={(2!)}^{-1/2}
\chi_{\varsigma}(\overrightarrow{M})\sum_{P}\delta_P \widehat{P}
\phi_{m}(\overrightarrow{x^*_1})\phi_{n}(\overrightarrow{x^*_2})$
where $\widehat{P}$ is one of the $2!$ permutation operators.
$\delta_P$ equals $+1$ for symmetric and $-1$ for anti-symmetric
case. The quantum number $m$ [e.g. $m=(n_r,l,s_z)$] and $n$ (for
electron two) characterize the two electron configuration $\{m,n\}$,
and the operator $\widehat{P}$ acts on the order of these single
electron configurations.

Integrating the right-hand side of Eq. (\ref{e:Aij}) over
$\overrightarrow{r_1}$ and $\overrightarrow{r_2}$, we end up with:
\begin{align}\label{e:Aif_final}
A_{ij}=(2!)^{-1} \delta_{M_z^i,M_z^{j}} \sum_{P}\sum_{P^{'}}\delta_P
\delta_{P^{'}}\widehat{P}\widehat{P^{'}}\{\sum_{\alpha}^{N_C}\sum_{\alpha^{'}}^{N_C}\delta_{n,n^{'}}
c_{\alpha}^{*}c_{\alpha^{'}}
\nonumber\\
\times\delta_{{s_{1z}^{\alpha}}{s_{1z}^{\alpha^{'}}}} A_{ij}^{\alpha
\alpha^{'}}+\sum_{\beta}\sum_{\beta^{'}}\delta_{m,m^{'}}
c_{\beta}^{*}c_{\beta^{'}}
\delta_{{s_{2z}^{\beta}}{s_{2z}^{\beta^{'}}}} A_{ij}^{\beta
\beta^{'}} \},
\end{align}
where $\alpha$ ($\in P$) and $\alpha^{'}$ ($\in P^{'}$) stand for
one electron and $\beta$ ($\in P$) and $\beta^{'}$ ($\in P^{'}$) for
the other electron. The prime indicates the final state $j$. The
matrix element $A_{ij}^{\alpha
\alpha^{'}}=\delta_{n_{r_{\alpha^{'}}},n_{r_{\alpha}}}\delta_{l_{\alpha^{'}},l_{\alpha}\pm1}l_H
\sqrt{n_{r_{\alpha}}+|l_{\alpha}|+1}
-\delta_{n_{r_{\alpha^{'}}},n_{r_{\alpha}}+1}\delta_{l_{\alpha^{'}},l_{\alpha}\pm1}
(1-\delta_{l_{\alpha},0})l_H\sqrt{n_{r_{\alpha}}+1}$ was calculated
before and was given by Eq. (18) of Ref.\cite{Geerinckx}.

We use the following Lorentzian broadened formula for the different
CR peaks:
\begin{equation}\label{e:sigmaE}
\sigma_i(E)=\sum_j \frac{\Gamma_{ij}}{\pi} \cdot\frac
{f_{ij}}{(E-E_{ij})^2+\Gamma_{ij}^2},
\end{equation}
where $i$ and $j$ refer to the initial and final states,
respectively. In the following results we concentrate on the
transitions where the initial state is the ground state (GS) and we
will identify $\sigma_1(E)$ by $\sigma(E)$. $\Gamma_{ij}$ is the
broadening parameter that is taken to be about $0.1\div1$ meV in our
numerical calculations.

We first calculate the oscillator strength (OS) for the case without
and with the presence of the Mn-ion taking into account all allowed
transitions of the two electrons. The system with the Mn-ion located
at the center of the dot has the energy spectrum as shown in
Fig.~\ref{spectrum_cen_lamb1_Ne=2} when the effective Coulomb
interaction strength is $\lambda_C=0.5$. Many crossings and
anti-crossings are found in the spectrum that were not present in
the case without a Mn-ion. The anti-crossings, which are a
consequence of intermixing of higher quantum states due to the
presence of the Mn-ion, lead to energy gaps between the levels, and
result in unusual behaviors in the cyclotron resonance spectrum.

For $N_e=2$, the pure ferromagnetic (FM) phase, where the two electrons
have spins parallel to the spin of the Mn-ion, does not exist. This
is opposite to the case for $N_e=1$ which is a consequence of the
closed $s$ orbital for $N_e=2$. This results in zero diagonal
elements for the e-Mn exchange matrix. The neighboring off-diagonal
elements that describe the spin exchange of the electrons with the
Mn-ion of the configurations with different $S_z$ and $M_z$ but
satisfying $S_z+M_z=$const., which are in general very small, now
turn to be the main contributions to the exchange interaction
energy. Therefore, a $\it{weakened}$ ``FM" state is still found with a
total spin slightly larger than zero. The magnitude of these
off-diagonal contributions is small and depends on the position of the magnetic
ion. Moving the Mn-ion to other positions inside the QD when the system
is still in the ``FM" phase can affect the structure of the energy
spectrum of the system. With increasing the magnetic field, the
system will transit to a phase where the two electrons will have spin-up and
are antiferromagnetically attracted to the Mn-ion. This phase is called
antiferromagnetic (AFM). The FM-AFM transition can be seen from
the inset of Fig.~\ref{spectrum_cen_lamb1_Ne=2} through the
anti-crossing between the GS and the second level at
$\Omega_c=1.81$. The Mn-ion is always ``frozen" with spin projection
$-5/2$ in case of nonzero magnetic field.
\begin{figure}[btp]
\begin{center}
\vspace*{-0.5cm}
\includegraphics*[height=8.8cm]{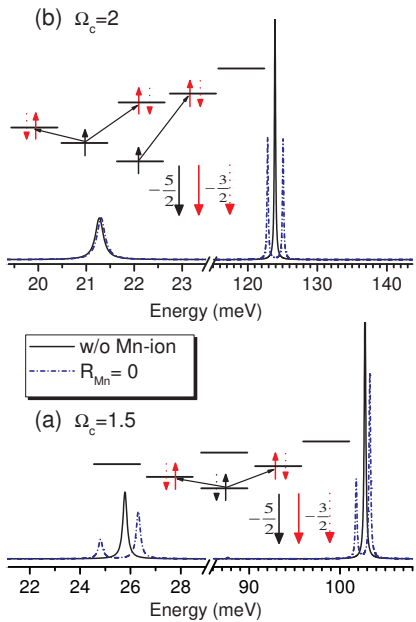}
\end{center}
\vspace{-0.5cm} \caption{(Color online) $\sigma(E)$ vs. energy for
two different magnetic fields where the system is in the (a) FM and
in the (b) AFM state for a two-electron QD with (blue dash-dotted
curve) and without (black solid curve) the presence of a Mn-ion for
$\lambda_C=0.5$. The Mn-ion is positioned in the center of the QD.
Two schematic plots in (a) and (b) describe the corresponding
transitions in which black is for GS and red for excited state;
solid is for main contribution and dotted for minor contribution to
the total OS.} \label{sigma_UC+cen_Ne=2}
\end{figure}

Figs.~\ref{sigma_UC+cen_Ne=2}(a) and ~\ref{sigma_UC+cen_Ne=2}(b) are
the magneto-optical absorption spectrum $\sigma(E)$ for two values
of the magnetic field $\Omega_c=1.5$ (FM) and $\Omega_c=2$ (AFM),
respectively, for the system in case without (black solid) and with
(blue dash-dot) the presence of the Mn-ion (at the center of the
dot). Two schematic diagrams describe possible electron transitions
with OS exceeding $0.1\%$ of the total OS taking place within the
$s$, $p$, and $d$ shells. When the system is in the FM phase, the
main transitions are such that the final state has the quantum
numbers $(n_r,L_z,S_z,M_z)=(0,1,0,-5/2)$ (the second peak) or
$(0,-1,0,-5/2)$ (the fourth peak) as the major configuration and
$(0,1,-1,-3/2)$ (the first peak) or $(0,-1,-1,-3/2)$ (the third
peak) as the minor configuration, respectively. We
recall\cite{NTTNguyen1} that the number of CR lines in the case of a
single electron quantum dot doped with a Mn-ion when the system is
in the FM state is two and the e (spin down)-Mn-ion (spin down)
interaction affects the CR spectrum in this state through shifting
the cyclotron energy and/or the presence of crossings. The spin
exchange interaction becomes stronger when the system is in the AFM
state (electron and Mn-ion spin antiparallel) resulting in the
presence of more CR lines. However, the major CR lines in this case
are two. Let us go back to the current system when it is in the AFM
state where we find three major CR lines as can be seen in
Fig.~\ref{sigma_UC+cen_Ne=2}(b). The first peak is due to an
electron transition from the $p^{+}$ [$(n_r,l)=(0,1)$] orbital to
the $d^{+}$ (0,2) orbital corresponding to the resonance from the GS
with major configuration $(0,1,1,-5/2)$ to the final state with
major configuration $(0,2,1,-5/2)$. For the other two pronounced
peaks, the one that appears at the smaller transition energy is for
the $s$-electron transition to the $p^{-}$ ($0,-1$) and its
oscillator strength is slightly larger than the other peak that
stands for the electron transition from the $p^{+}$ (0,1) orbital to
the $d^{0}$ ($1,0$) orbital. Their final states have as dominant
configurations $(0,0,1,-5/2)$ and $(1,0,1,-5/2)$, respectively. The
CR transitions that become allowed due to spin exchange have a much
smaller oscillator strength than the major transitions and can be
neglected. For the case without a Mn-ion, the system transits from
the state $(n_r,L_z,S_z)=(0,0,0)$ to the state $(0,1,1)$. Note that
if one keeps increasing the magnetic field, the electrons will
occupy higher quantum states resulting in the GS wave function
having quantum numbers e.g. $(0,3,1)$ or $(1,1,1)$, and so on.
Consequently, the number of possible CR lines increases.

Here,
we will discuss in more detail the transitions resulting from the
$(0,1,1)$ GS. Note that only two of the three allowed CR lines are
observable, namely, the $p^{+}$-electron to the $d^{+}$ [the first
solid peak in Fig.~\ref{sigma_UC+cen_Ne=2}(b)] and the $s$-electron
and the $p^{+}$-electron to, respectively, the $p^{-}$ and the
$d^{0}$ orbital that have the same OS and transition energy [the
second solid peak in Fig.~\ref{sigma_UC+cen_Ne=2}(b)]. These two
coinciding CR lines split in the presence of the Mn-ion as can be
seen by the dash-dotted curve where the system is in the AFM state.
We note that for $\Omega_c=2$ [Fig.~\ref{sigma_UC+cen_Ne=2}(b)] the
first peak corresponding to the transition of the $p^{+}$-electron
to the $d^{+}$ orbital has almost the same transition energy in case
without and with the Mn-ion. This is due to the zero exchange
interaction energy between the $s$ and the $p$ orbitals when the
Mn-ion is located at the center of the dot. The energy difference
between these two peaks and the number of spin-exchange CR lines
will increase as one moves the Mn-ion away from the center of the
QD. We will come back to this point in a later discussion.
\begin{figure}[btp]
\begin{center}
\vspace*{-0.5cm}
\includegraphics*[width=8.5cm]{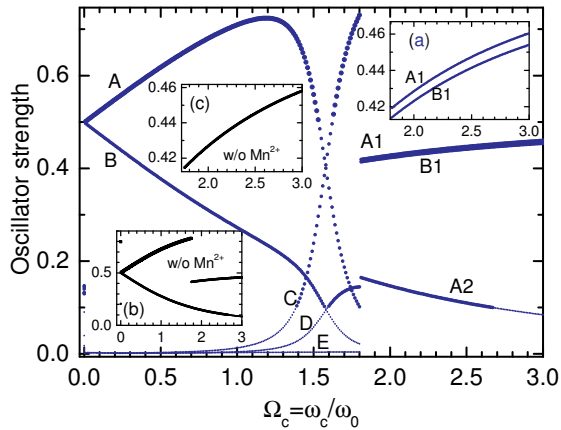}
\end{center}
\vspace{-0.5cm} \caption{(Color online) Oscillator strength of a
two-electron QD with the Mn-ion located at the center of the QD, (a)
magnification of the region of $1.8\le\Omega_c\le3$ where the system
transits to the AFM phase. (b) the OS of the same QD without a
Mn-ion and (c) the magnification of the same region as in (a) but
for the same situation as in (b). The thickness of the curve is
proportional to the value of the OS. A, B, C, D, E labels the $5$
branches when the system is in the FM and A1, A2, and B1 labels the
branches when the system is in the AFM phase.}
\label{OS_cen_Ne=2_value}
\end{figure}
\begin{figure}[btp]
\begin{center}
\vspace*{-0.5cm}
\includegraphics*[width=8.3cm]{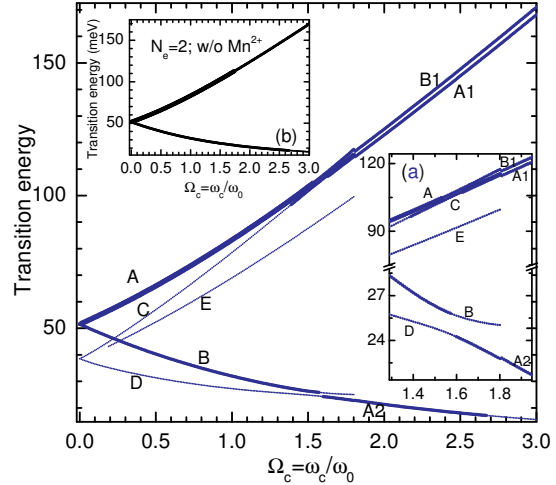}
\end{center}
\vspace{-0.5cm} \caption{(Color online) Transition energy
corresponding to the OS of Fig.~\ref{OS_cen_Ne=2_value} and for the
case without a Mn-ion [inset (b)]. (a) Magnification of the
transition energy of the main plot for the magnetic field region
$1.3\le\Omega_c\le2.0$. The thickness of the curve is proportional
to the respective OS value and the labels corresponding to the
transition energy branches of the respective OS ones plotted in
Fig.~\ref{OS_cen_Ne=2_value}.} \label{ETRAN_cen_Ne=2_value}
\end{figure}

The OS and absorption energy spectrum of the above case over a wide
range of magnetic field are found in Fig.~\ref{OS_cen_Ne=2_value}
and Fig.~\ref{ETRAN_cen_Ne=2_value}, respectively. For the case
without a Mn-ion, these quantities are, respectively, reviewed in
the insets (b) and (c) of Fig.~\ref{OS_cen_Ne=2_value} and in the
inset (b) of Fig.~\ref{ETRAN_cen_Ne=2_value}.
Fig.~\ref{ETRAN_cen_Ne=2_value}(b) is identical to the case of
one-electron QD due to the $N_e$-independence of the absorption
spectrum for parabolic confinement without a Mn-ion. The thickness
of the transition absorption energy curves plotted in
Fig.~\ref{ETRAN_cen_Ne=2_value} is proportional to the OS whose
values are plotted in Fig.~\ref{OS_cen_Ne=2_value}. As we can see
from these two plots, the CR transitions for the case when the
Mn-ion is present is very different from the case without a Mn-ion.
For the magnetic field region where the system is in the FM phase
$\Omega_c\le1.81$, the differences are significant which can be seen
from the presence of the two crossings in the OS in
Fig.~\ref{OS_cen_Ne=2_value} that correspond to the two
anti-crossings in the transition energy in
Fig.~\ref{ETRAN_cen_Ne=2_value}. At the FM-AFM transition point,
there are discontinuities. For the AFM region, the difference is in
the separation of the two lines that was not present in the case
without a Mn-ion as shown, respectively, in
Figs.~\ref{OS_cen_Ne=2_value}(a), ~\ref{ETRAN_cen_Ne=2_value}(a) and
Figs.~\ref{OS_cen_Ne=2_value}(b), ~\ref{ETRAN_cen_Ne=2_value}(b).
Note that the discontinuity in the OS of
Fig.~\ref{OS_cen_Ne=2_value}(b) is due to the fact that the CR
spectrum transits from two to three lines, where the last two CR
lines have the same transition energy (they are the transitions $s
\rightarrow p^{-}$ and $p^{+} \rightarrow d^{0}$). If these OS are
added, there would be no discontinuity in
Fig.~\ref{OS_cen_Ne=2_value}(b). Let us first discuss the branches
of the OS for the region $\Omega_c<1.4$ (before the first crossing)
- which we name region I. In this region, the system is in the FM
phase. The two electrons mostly stay in the $s$ shell with
antiparallel spins. The transitions are from the GS to the two $p$
orbitals with high OS . The latter are the two higher branches in
the OS curve plotted in Fig.~\ref{OS_cen_Ne=2_value}. The final
states, which now mix with several quantum states, include,
respectively, $(0,-1,0,-5/2)$ (higher) or $(0,1,0,-5/2)$ as their
main contribution. We see also two lower branches; that are the
curves corresponding to the electron transitions from the GS to the
two $p$ orbitals but with different spin states. The final states
are now such that the electron that is excited to a higher state
will flip its spin which is compensated by a change of the Mn-ion
spin. We found that the final states corresponding to these CR lines
have, respectively, the quantum numbers $(0,-1,-1,-3/2)$ (higher)
and ($0,1,-1,-3/2$) as their major contribution. This is understood
via the e-Mn-ion spin-spin exchange interaction. These two curves
have a smaller OS than the above two due to the small contribution
of the configuration $(0,0,-1,-3/2)$ in the GS while the major one
is $(0,0,0,-5/2)$.

Now we will discuss the results for the higher magnetic field
region. At the first crossing, $\Omega_c=1.45$, we observe the
exchange between two branches as seen in
Fig.~\ref{OS_cen_Ne=2_value}. The higher branch of the two lower
branches is higher in energy than the lower branch of the two higher
branches in region I of the OS plot. This is the case up to the next
two crossings at $\Omega_c=1.58$. The region
$1.45\le\Omega_c\le1.58$ is named region II. Within this magnetic
field region, the final state which has the configuration
$(0,-1,-1,-3/2)$ as its major contribution to the OS becomes larger
in OS as compared to the state which has the configuration
$(0,1,0,-5/2)$. At $\Omega_c=1.58$ we find two crossings with an
exchange of OS. This is a consequence of the fact that the energy
spectrum exhibits crossings (and anti-crossings) of energy levels.
Next we consider the AFM phase, $\Omega_c\ge1.81$, and focus on the
two higher branches that stay very close to each other. Remember
that for the case without a Mn-ion, these branches are degenerate
[as can be seen from Fig.~\ref{OS_cen_Ne=2_value}(c) and
Fig.~\ref{ETRAN_cen_Ne=2_value}(b)]. In the presence of the Mn-ion,
this degeneracy is lifted and and their OS differ about $1\%$ of the
total OS and their energy differs by $3-4$ meV or about $6.8\%$ of
the confinement energy $\hbar\omega_0$. They correspond to
transitions of the spin-up electron from the $s$ shell to the
$p^{-}$ orbital (lower branch) and of the spin-up electron from the
$p^{+}$ orbital to the $d^{0}$ orbital as discussed before in
Fig.~\ref{sigma_UC+cen_Ne=2}, respectively. For these transitions,
the Mn-ion spin with $M_z=-5/2$ is unaltered. The other lower branch
corresponds to the transition of the $p^{+}$-electron to the $d^{+}$
orbital. The anti-crossings in the transition energy are illustrated
in Fig.~\ref{ETRAN_cen_Ne=2_value}(a) which magnifies the region
$1.3\le\Omega_c\le2.0$.
\begin{figure}[btp]
\begin{center}
\vspace*{-0.5cm}
\includegraphics*[width=8.5cm]{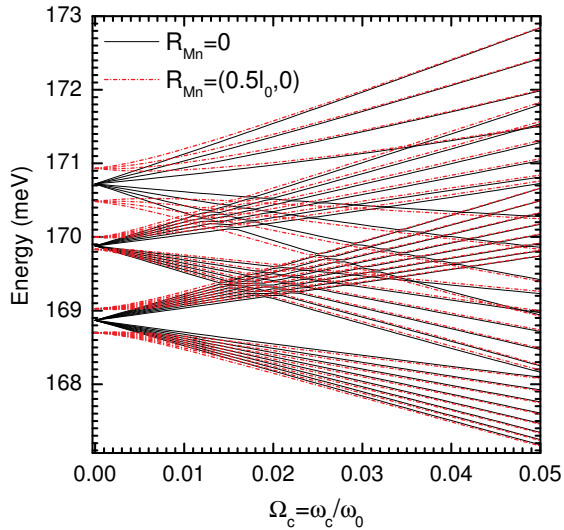}
\end{center}
\vspace{-0.5cm} \caption{(Color online) $36$ energy levels (from
level $7$ to level $42$ that are grouped into 8-8-6-6-4-4 lines) of
a two-electron QD with the Mn-ion located at the center or at
$(0.5l_0,0)$ of the QD with the Coulomb interaction strength
$\lambda_C=0.5$.}\label{Midspec_cen+0.5l0_lamb1_Ne=2}
\end{figure}

\subsection{Mn-ion position dependence of
the intra-band excitation spectrum} Moving the Mn-ion to another
location inside the QD affects the energy spectrum as shown in
Fig.~\ref{Midspec_cen+0.5l0_lamb1_Ne=2} where we focus on the small
magnetic field region. The black solid lines and the red dashed
lines are the $36$ energy levels starting from level $7$ for the
cases that the Mn-ion is located at the center and at $(0.5l_0,0)$,
respectively. We first notice that the $B=0$ GS energy in both cases
has the same degeneracy (i.e. $6$). For the higher energy levels
different degeneracies are found for $R_{Mn}=0$ we have $16-12-8$
while for the other case we have $8-8-6-6-4-4$ fold degeneracy. When
applying a magnetic field, these degeneracies are further lifted due
to the Zeeman effect. We note that e.g. for the two
eightfold-degenerate levels [in the case that the Mn-ion is located
at $(0.5l_0,0)$-the red dash-dotted curves] or the two
$16$-fold-degenerate-levels (in the case that the Mn-ion is located
at the center of the dot-the black solid curves), as the magnetic
field increases, the lower energy level corresponds to states with
positive total azimuthal quantum number, in this case $+1$ - one
electron in the $s$ shell and the other in the $p^{+}$ orbital and
the higher energy level has negative total azimuthal quantum number
$-1$ - one electron in the $s$ shell and the other in the $p^{-}$
orbital, etc. The degeneracy of the higher energy levels for $B=0$,
i.e. $8$, $6$, or $4$, etc, comes from the ferromagnetic (e.g.
$\overrightarrow{M}+\overrightarrow{S}=\overrightarrow{7/2}$ for the
case of eightfold degeneracy) or antiferromagnetic (e.g.
$\overrightarrow{M}+\overrightarrow{S}=\overrightarrow{5/2}$ for the
case of sixfold degeneracy) coupling of the two electrons with the
Mn-ion. Note that all the $36$ energy levels in
Fig.~\ref{Midspec_cen+0.5l0_lamb1_Ne=2} refer to the $s$ and $p$
orbitals. At higher magnetic fields, the energy spectrum exhibits
further differences when $R_{Mn}$ is varied in e.g. the number and
the positions of the crossings and anti-crossings, the energy gaps
of the anti-crossings, etc.
\begin{figure}[btp]
\begin{center}
\vspace*{-0.5cm}
\includegraphics*[width=9.0cm]{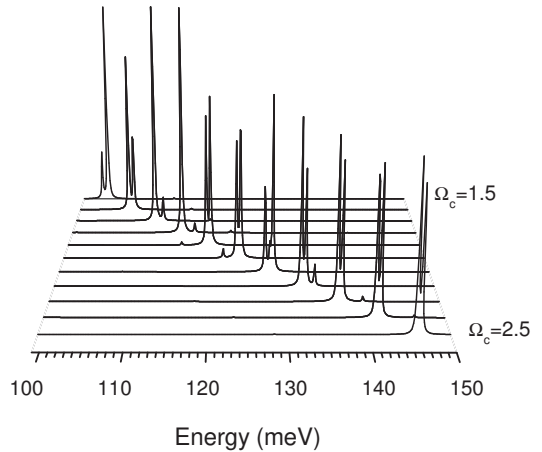}
\end{center}
\vspace{-0.5cm} \caption{(Color online) Magneto-optical absorption
spectrum focused on the two higher transitions for the same QD as
plotted in Fig.~\ref{ETRAN_cen_Ne=2_value} for the case the Mn-ion
is located at $(0.5l_0,0)$. The magnetic field range is (1.5,2.5)
with steps of $0.1$.} \label{ETRAN_0.5l0_lamb1_3D}
\end{figure}

As we discussed above, the system that consists of two electrons and
a single Mn-ion does not exhibit a very clear FM phase since the
state where the two electrons stay in the $s$ shell with
antiparallel spins minimizes the GS. The energy gained by the direct
antiferromagnetic coupling of the Mn-ion to the spin-up electron is
balanced by the ferromagnetic exchange with the other spin-down
electron. This holds as long as the two electrons occupy the $s$
orbital. Therefore, the FM-AFM transition magnetic field is almost not
changed when we vary the position of the Mn-ion. For example, the
FM-AFM takes place at $\Omega_c=1.81$ for the cases that the Mn-ion
is located at the center and at $(0.5l_0,0)$ and at $\Omega_c=1.79$
for the case the Mn-ion is located at $(l_0,0)$. To see the Mn-ion
position dependence on the CR-spectrum, we will focus on the high
magnetic field region where the system is in the AFM phase with both
electrons having spin up and one of them accommodating a higher
orbital e.g. the $p^{+}$ orbital. The allowed electron transitions
now are the three lines that start from the GS to the $p^{-}$ (for
the $s$-electron) and to the $d^{+}$ and $d^{0}$ orbitals (for the
$p$-electron). Note that for the case that the Mn-ion is located at
the center of the dot, the exchange interactions between the $s$ and
$p$ orbitals are zero while these terms increase as the Mn-ion is
moved away from the center e.g. to $(0.5l_0,0)$ or $(l_0,0)$.
Consequently, the behavior of the OS changes. The first transition
that corresponds to the transition of the $p^{+}$-electron to the
$d^{+}$ orbital is shifted. The other two transitions that stay very
close in energy and correspond to the other transitions of the
$s$-electron and the $p^{+}$-electron to the $p^{-}$ and $d^{0}$
orbitals [the solid lines at the right hand side in the diagram in
Fig.~\ref{sigma_UC+cen_Ne=2}(b)], respectively, are most separated
for the case the Mn-ion is located at the center of the QD.
The reason is that for $R_{Mn}=0$, only the exchange interactions between the
$s$ and the $d^{0}$ orbitals are nonzero (and equal) within the
involved shells $s$, $p$, and $d$. This means that the final
state with the configuration of the $s$-electron and the
$d^{0}$-electron as their main contribution to the wave function is more
enhanced in energy. This difference leads to a larger
separation between the two CR lines for $R_{Mn}=0$. Qualitatively,
this separation is about $3-4$ meV for $R_{Mn}=0$ and $0.9-1$ meV for
$R_{Mn}=0.5l_0$, $0.5$ meV for $R_{Mn}=l_0$ and almost zero for
$R_{Mn}=2l_0$. In the last case the problem converts to the problem
without a Mn-ion. The two transitions merge to a single one as seen
before [black curve in Fig.~\ref{sigma_UC+cen_Ne=2}(b)].
\begin{figure}[btp]
\begin{center}
\vspace*{-0.5cm}
\includegraphics*[width=8.8cm]{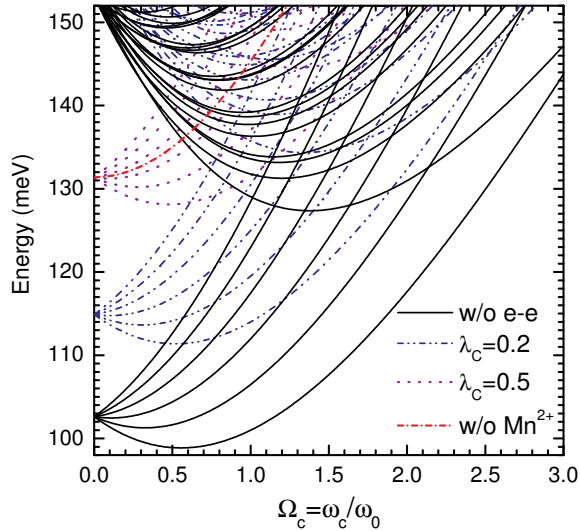}
\end{center}
\vspace{-0.5cm} \caption{(Color online) Energy spectrum plotted for
the first $24$ levels for several different Coulomb interaction
strengths: without e-e interaction, $\lambda_C=0.2$, $0.5$. Red
dash-dotted curve is the energy of the same QD ($\lambda_C=0.5$) without a Mn-ion for
reference. The energy spectrum for the case $\lambda_C=0.2$ is
scaled to the case of $\lambda_C=0.5$.}
\label{spectrum_twolamb_Ne=2}
\end{figure}

We complete this subsection by investigating the absorption spectrum
of the system for $R_{Mn}=0.5l_0$ in a range of magnetic field
$\Omega_c=(1.5,2.5)$ that includes the FM-AFM transition [see
Fig.~\ref{ETRAN_0.5l0_lamb1_3D}]. We focus our discussions on the
two transitions discussed in the previous paragraph that correspond
to the right hand side CR lines in Fig.~\ref{sigma_UC+cen_Ne=2}(b).
As the system transits to the AFM phase, the ``correlations" between
the $s$ and $p$, $s$ and $d$, $p$ and $d$, etc, become nonzero for
almost all values of the coupling strength. Consequently, the final
states of the two transitions gain energy from the exchange
interaction part. Their difference reduces leading to the fact that
these two peaks stay closer in energy as compared to the case that
the Mn-ion is located at the center of the dot. However, the
stronger peak (at smaller transition energy) which corresponds to
the transition of the $s$-electron when the system is in the AFM
phase can become more pronounced or smaller than the other
transition as illustrated in Fig.~\ref{ETRAN_0.5l0_lamb1_3D}. As the
system transits to the AFM phase ($\Omega_c\ge1.9$), we see that
within the magnetic field range $\Omega_c=1.9\div2.1$ the peak that
corresponds to the transition of the $p^{+}$ electron to the $d^{0}$
orbital is higher (having a larger OS) as compared to the other peak
that corresponds to the transition of the $s$-electron to the
$p^{-}$ orbital. Within the magnetic field where
$\Omega_c=2.2\div2.3$ the OS of the two peaks exchange. This
exchange happens again at $\Omega_c=2.4$ and we obtain the last
exchange in this figure for $\Omega_c=2.5$. Each time that there is
an OS exchange between these two peaks there is a crossing in the OS
(and an anti-crossing) in the transition energy. When the $Mn^{2+}$
is displaced to $(0.5l_0,0)$, the difference in the OS value between
these two peaks is now larger, but the difference in energy is
smaller. Note that also smaller peaks appear next to the main ones.
The small peak for $\Omega_c=1.5$ is a transition from the GS with
dominant configuration $(0,0,0,-5/2)$ which has $(0,0,-1,-3/2)$ as a
minor contribution, to the state with the configuration
$(0,-1,-1,-3/2)$ as its main contribution to the wave function. For
$\Omega_c=1.9$ the small peak is due to the transition to the state
with the configuration $(0,-1,-1,-3/2)$ as its main contribution.
\begin{figure}[btp]
\begin{center}
\vspace*{-0.5cm}
\includegraphics*[width=8.5cm]{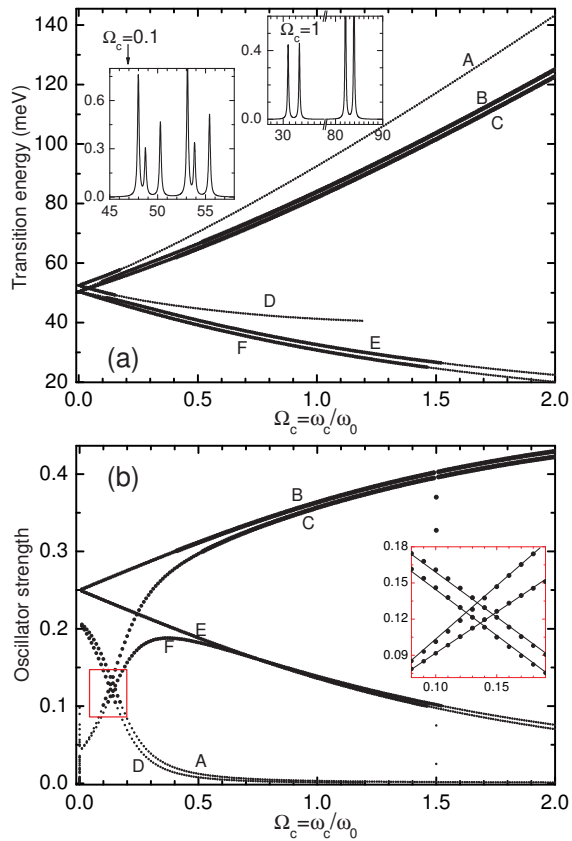}
\end{center}
\vspace{-0.5cm} \caption{(Color online) Transition energy (a) and OS
(b) of a two-electron QD without e-e interaction for the case the
Mn-ion is at the center of the dot. The thickness of the curve is
proportional to the OS value (up to $\Omega_c=2$, there is still no
FM-AFM transition happening). The two insets in (a) are the
$\sigma(E)$ plotted for two magnetic fields: $\Omega_c=0.1$ and $1$.
A, B, C, D, E, and F label the six branches of the transition energy
in (a) and the respective OS branches in (b). The inset in (b) is a
magnification of the red rectangular where we use four lines to
connect the data points of the four corresponding CR lines.}
\label{ETRAN_noCou+UC_Ne=2}
\end{figure}

\subsection{Electron-electron interaction effect}
Here we investigate how the strength of the e-e interaction influences
the CR absorption spectrum. Its influence on the first $24$ energy levels
is plotted in Fig.~\ref{spectrum_twolamb_Ne=2} in case without e-e interaction
(solid black curves) and two different $\lambda_C$ [$0.2$ (blue
dash-dot-dotted curves) and $0.5$ (violet dash curves)]. Crossings and
anti-crossings of these energy levels occur at different magnetic
fields (the smaller $\lambda_C$ the larger the crossing field $\Omega_c$).

Lets turn to the transition energy and for reference purposes we
turn off the Coulomb interaction strength (see
Fig.~\ref{ETRAN_noCou+UC_Ne=2}). Without the Coulomb interaction,
the FM-AFM transition takes place at a larger magnetic field
$\Omega_c=3.44$ as compared to $\Omega_c=1.81$ for the case with e-e
interaction (see Fig.~\ref{ETRAN_cen_Ne=2_value} and
Ref.\cite{NTTNguyen2}). There are generally six CR lines (as
compared to four for the case with e-e interaction) with OS
exceeding $0.1\%$ of the total OS
[Fig.~\ref{ETRAN_noCou+UC_Ne=2}(b)]. The main difference is that the
dominant CR lines [conventional transitions e.g. from the GS with
the quantum state $(0,0,0,-5/2)$ as the main configuration to the
final state with the quantum state $(0,\pm1,0,-5/2)$] do not have a
much larger OS than the secondary transitions - the transitions
appearing as a consequence of the e-Mn-ion exchange interactions, as
can be seen from Figs.~\ref{OS_cen_Ne=2_value} and
~\ref{ETRAN_noCou+UC_Ne=2}(b) ($\le 15 \%$ of the total OS). These
six lines correspond to transitions of the electrons from the GS
containing the major quantum state $(n_r,L_z,S_z,M_z)$, to the six
final states that contain their dominant contributions to the wave
functions as combinations of the following four quantum states:
$(n_r,L_z+1,S_z,M_z)$, $(n_r,L_z+1,S_z-1,M_z+1)$,
$(n_r,L_z-1,S_z,M_z+1)$, and $(n_r,L_z-1,S_z-1,M_z+1)$. Depending on
the value of the applied magnetic field, the relative contributions
of these states to the wave function will change leading to
crossings as seen at $\Omega_c=0.13$, $0.14$ in the inset of
Fig.~\ref{ETRAN_noCou+UC_Ne=2}(b). If we increase the magnetic field
further the CR spectrum collapses into four CR lines. The lines that
are the results of the exchange e-Mn-ion interaction (C and F)
become close in energy to the ``conventional" lines (B and E). This
behavior is not seen in case of interacting electrons. The reason is
that when the Coulomb interaction is turned on, see
Fig.~\ref{ETRAN_cen_Ne=2_value}, the e-Mn spin-spin interaction is
weakened by the Coulomb repulsion. Note that the two additional
peaks in the CR spectrum are a consequence of the breaking of the
selection rule by the e-Mn exchange interaction.
\begin{figure*}[btp]
\begin{center}
\vspace*{-0.5cm}
\includegraphics*[width=14.0cm]{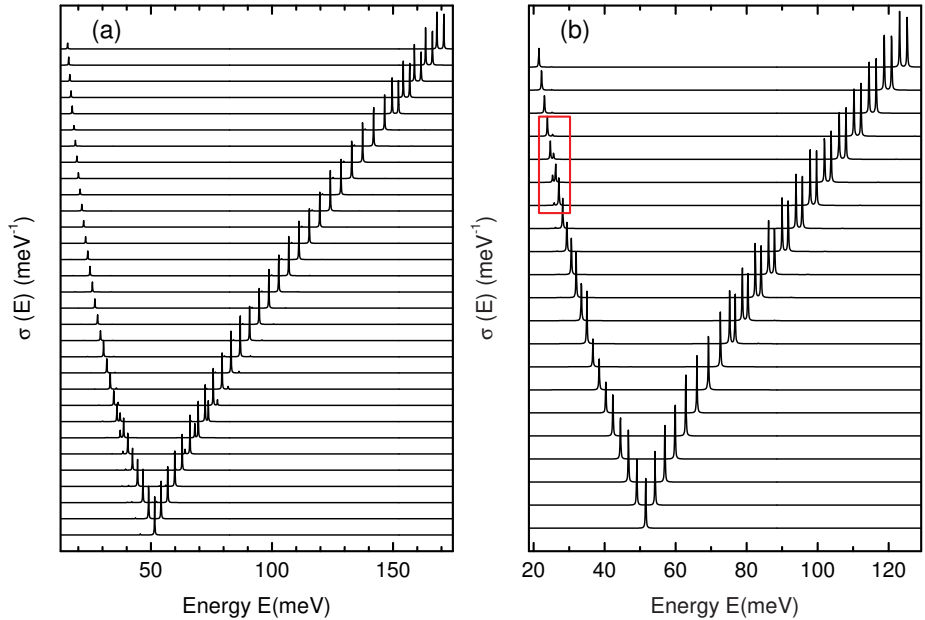}
\end{center}
\vspace{-0.5cm} \caption{(Color online) Magneto-optical absorption
spectrum scaled to the case of $\lambda_C=0.5$ obtained for the case
the Mn-ion is located at the center of the QD for $\lambda_C=0.2$
(a) and $\lambda_C=2$ (b).
Results are shown in the magnetic field range $(0,3)$ (a) and $(0,2)$ (b) with steps of
$0.1$.} \label{ETRAN_cen_lamb0.4+4_Ne=2}
\end{figure*}

When comparing the low magnetic field behavior
(see Figs.~\ref{OS_cen_Ne=2_value} and ~\ref{ETRAN_cen_Ne=2_value}) of the CR spectrum we note
a very different behavior when we turn off the e-e interaction (see Fig.~\ref{ETRAN_noCou+UC_Ne=2}).
To understand this different behavior we plot in Fig.~\ref{ETRAN_cen_lamb0.4+4_Ne=2}(a) the magneto-optical absorption
spectrum for the system with a Mn-ion located at the
center of the dot and $\lambda_C=0.2$ for the magnetic field range
$0<\Omega_c<3$. As compared to the system studied in
Figs.~\ref{OS_cen_Ne=2_value} and ~\ref{ETRAN_cen_Ne=2_value} with
$\lambda_c=0.5$: first, the AFM phase takes place at a larger
magnetic field ($\Omega_c=2.53$ as compared to $1.81$), second, the
low OS branches survive up to higher fields. The FM-AFM transition
occurs at $\Omega_c=1.24$ and $0.96$, for $\lambda_C=1$ and $1.5$,
respectively. As the e-e interaction strength increases, e.g. from
$\lambda_C=0.2$ [see Fig.~\ref{ETRAN_cen_lamb0.4+4_Ne=2}(a)] to
$\lambda_C=2$ [see Fig.~\ref{ETRAN_cen_lamb0.4+4_Ne=2}(b)], the low OS
branches that appear due to the spin exchange of the electrons and
the Mn-ion contribute less to the total OS. In
Fig.~\ref{ETRAN_cen_lamb0.4+4_Ne=2}(a), the contribution of the exchange
terms to the total OS stays appreciable within the magnetic field
range $(0,1.7)$ (FM). The low OS branches appear on both ``sides" of
the transitions of the $s$-electrons: to the $p^{+}$ (left - lower
transition energy) and to the $p^{-}$ (right - higher transition
energy) orbitals [see the lower peaks of
Fig.~\ref{ETRAN_cen_lamb0.4+4_Ne=2}(a)], meaning that the final states
with the configurations having $(0,0,S_z+M_z=-5/2)$ as the main
contribution have nonzero OS. While these low OS branches appear only
on the ``side" of the $p^{+}$-electron resonating to the $d^{+}$
orbital within the magnetic field region (1.4,1.7) (AFM), see the red
rectangular in Fig.~\ref{ETRAN_cen_lamb0.4+4_Ne=2}(b). It means that only
the final state with the main quantum configuration as
$(0,2,0,-3/2)$ has a nonzero contribution (as a minor) to the
oscillator strength, while there is no gain from the exchange
interactions during the transitions of the $p^{+}$-electron and the
$s$-electron to the $d^{0}$ and $p^{-}$ orbitals, respectively.
However, in any case, the exchange interactions always separate
these two transitions in energy (about $6.8\%$ of $\hbar\omega_0$
for $R_{Mn}=0$), as can be observed in
Figs.~\ref{ETRAN_cen_lamb0.4+4_Ne=2}(a) and ~\ref{ETRAN_cen_lamb0.4+4_Ne=2}(b),
for magnetic field $\Omega_c\ge2.53$ and $\Omega_c\ge0.79$ for
$\lambda_C=0.2$ and $2$, respectively.

\begin{figure}[btp]
\begin{center}
\vspace*{-0.5cm}
\includegraphics*[width=8.5cm]{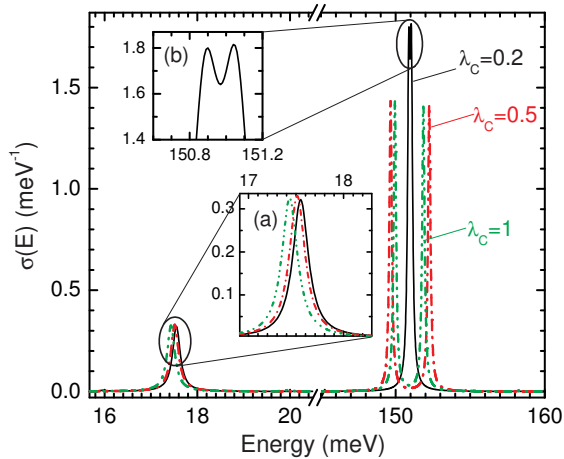}
\end{center}
\vspace{-0.5cm} \caption{(Color online) Magneto-optical absorption
spectrum scaled to the case of $\lambda_C=0.5$ obtained for the case
the Mn-ion is located at $(0.5l_0,0)$ for three different Coulomb
interaction strengths $\lambda_C=0.2$, $0.5$, and $1$ and magnetic
field $\Omega_c=2.6$. (a) magnification of the first major peak. We
focused on the three main transitions in each case with inset (b)
the magnification to see the two upper peaks for the case of
$\lambda_C=0.2$. } \label{sigma_lamb_Imp0.5l0_Ne=2}
\end{figure}

Last, we study the combined effects of the position of the
Mn-ion and the Coulomb
interaction strength on the CR absorption spectrum. Due to the
exchange interactions between all included quantum orbitals, the
transitions of the $s$-electron to the $p^{-}$ orbital and of the
$p^{+}$-electron to the $d^{0}$ orbital have different energies
depending on the strength of the Coulomb interaction as can be seen
in Fig.~\ref{sigma_lamb_Imp0.5l0_Ne=2}. We plot $\sigma(E)$ as a
function of energy for three different $\lambda_C$
(=$0.2$, $0.5$ , and $1$) at $\Omega_c=2.6$ (AFM phase) with the
Mn-ion located at $(0.5l_0,0)=(13.2\AA,0)$. For the smallest
considered $\lambda_C$ (i.e. $0.2$) the two electrons are more
strongly confined and the location of the Mn-ion in this case is
almost out of the effective region of the electrons. Consequently,
the separation between the two transitions of the electrons in two
different orbitals becomes very small as can be seen from the black
solid curve in Fig.~\ref{sigma_lamb_Imp0.5l0_Ne=2} and its inset.
The first transition, the transition of the $p^{+}$-electron to the
$d^{+}$ orbital, slightly changes as $\lambda_C$ changes as shown in
Fig.~\ref{sigma_lamb_Imp0.5l0_Ne=2}(a).
\section{Summary and Conclusions}
We find that the electron transition energy in parabolic quantum dot
with a single Mn-ion depends on the strength of the e-e interaction
and the number of electrons in the QD. The e-e interaction shifts
and splits the two main branches (the upper-energy ones) of the
absorption spectrum at the FM-AFM transition magnetic field. The
strength of the e-e interaction changes the position of the low OS
branches significantly. Different branches appear with changing
Coulomb interaction strength.

In a many-electron quantum dot without a Mn-ion the CR spectrum consists of
two peaks due to Kohn's theorem. The reason for the breakdown of
Kohn's theorem is the presence of
the Mn-ion, which leads to spin exchange interaction with the
electrons of the QD. Moving this Mn-ion to different positions in
the QD corresponds to changing the magnitude of the exchange
interactions and leads to changes in the absorption energy spectrum.

In the FM region, the spin-spin e-Mn-ion interactions share their
contributions to the total OS with the two ``direct" transitions of
the two $s$-electrons in which the total $S_z$ is conserved. In the
AFM region, this leads to discontinuities in the OS and the
respective transition energy and separate the two transitions of the
$s$- and the $p^{+}$-electrons in energy. These two transitions address,
respectively, the $p^{-}$ and $d^{0}$ orbitals. At the FM-AFM
transition magnetic field, we observed that the OS exhibits
a discontinuity for all the CR peaks. When the system is in the
FM phase, the transition energy exhibits only minor differences as compared to the case
without a Mn-ion. When the system is in the AFM phase, however,
there are major differences in the absorption energy spectrum. The
separation of the two upper peaks in the CR spectrum depends on the
effective e-e interaction strength and the position of the Mn-ion.
The heights of these two upper peaks change with changing position
of the Mn-ion and their relative heights exchange OS at each
anti-crossing in the transition energy. The electrons
spin-spin exchange with the Mn-ion and the strength of it depends on
the position of the Mn-ion. Therefore, changing the Coulomb
interaction strength will affect the absorption energy spectrum by
e.g. separating or merging the two above electron transitions. When
the Mn-ion is moved away from the center of the dot, the exchange
interactions become nonzero resulting in additional CR peaks. The
number of these branches reduces as the Coulomb interaction strength
increases. When the e-e interaction is turned off, we found that
the transitions that become allowed due to the spin exchanges increase in
OS.

In short, in the FM phase (i.e. low magnetic field situation) the
number of CR lines is increased to four for $N_e=2$ which compares
to two lines when $N_e=1$. In the AFM phase (i.e. high magnetic
field) the number of CR lines with substantial OS is reduced to
three while for $N_e=1$ it was four of which only two had an OS
appreciable different from zero. These changes in the CR spectrum
should be observable in a CR experiment. The number of electrons in the quantum dot can be varied
through the application of a gate potential. In order to resolve the
CR lines that are very close to each other it is recommended to use
single dot spectroscopy.\cite{Besombes,Leger,Gall}
\section{Acknowledgments}
This work was supported by FWO-Vl (Flemish Science Foundation), the
EU Network of Excellence: SANDiE, the Brazilian science foundation
CNPq, and the Belgian Science Policy (IAP).



\begin{thebibliography}{99}
\bibitem{Furdyna} J.K. Furdyna, J. Appl. Phys. \textbf{64}, R29 (1988).
\bibitem{Besombes} L. Besombes, Y. L\'{e}ger, L. Maingault, D. Ferrand,
H. Mariette, and J. Cibert, Phys. Rev. Lett. \textbf{93}, 207403
(2004).
\bibitem{Leger} Y. L\'{e}ger, L. Besombes, J. Fern\'{a}ndez-Rossier,
L. Maingault, and H. Mariette, Phys. Rev. Lett. \textbf{97}, 107401
(2006); L. Maingault, L. Besombes, Y. L\'{e}ger, H. Mariette, and C.
Bougerol, Phys. Status Solidi C \textbf{3}, 3992 (2006).
\bibitem{Gall} C. Le Gall, L. Besombes, H. Boukari, R. Kolodka, J. Cibert,
and H. Mariette, Phys. Rev. Lett. \textbf{102}, 127402 (2009).
\bibitem{Govorov} A. O. Govorov, Phys. Rev. B \textbf{70}, 035321
(2004).
\bibitem{Schmidt} T. Schmidt, M. Scheibner, L. Worschech, A. Forchel,
 T. Slobodskyy, and L. W. Molenkamp, J. Appl. Phys. \textbf{100}, 123109 (2006).
\bibitem{Worjnar}  P. Wojnar, J. Suffczy\'{n}ski, K. Kowalik, A. Golnik,
G. Karczewski, and J. Kossut, Phys. Rev. B \textbf{75}, 155301
(2007).
\bibitem{Fernandez} J. Fern\'{a}ndez-Rossier and Ram\'{o}n Aguado, Phys. Rev. Lett. \textbf{98},
106805 (2007); J. Fern\'{a}ndez-Rossier, Phys. Rev. B \textbf{73 },
045301 (2006).
\bibitem{Chang} K. Chang, J.B. Xia, and F.M. Peeters,
Appl. Phys. Lett. \textbf{82}, 2661 (2003).
\bibitem{Glazov} M. M. Glazov, E. L. Ivchenko, L. Besombes, Y. L\'{e}ger, L. Maingault,
and H. Mariette, Phys. Rev. B \textbf{75} 205313 (2007).
\bibitem{Hawrylak} F. Qu and P. Hawrylak, Phys. Rev. Lett. \textbf{95}, 217206
(2005).
\bibitem{NTTNguyen2} N. T. T. Nguyen and F. M. Peeters, Phys. Rev. B \textbf{78}, 045321 (2008).
\bibitem{Govorov_review} A. O. Govorov, C. R. Physique \textbf{9}, 857 (2008).
\bibitem{Kohn-Luttinger} W. Kohn and J. M. Luttinger, Phys. Rev. \textbf{96},
529 (1954); J. M. Luttinger, Phys. Rev. \textbf{102}, 1030 (1956).
\bibitem{Kohn} W. Kohn, Phys. Rev. \textbf{123}, 1242 (1961).
\bibitem{Peeters} F. M. Peeters, Phys. Rev. B \textbf{42}, 1486 (1990).
\bibitem{Sikorski} Ch. Sikorski and U. Merkt, Surf. Sci. \textbf{229}, 282 (1990).
\bibitem{Geerinckx} F. Geerinckx, F. M. Peeters, and J. T.
Devreese, J. Appl. Phys. \textbf{68}, 3435 (1990).
\bibitem{Helle} M Helle, A Harju, and R M Nieminen, New J. Phys. \textbf{8}, 27
(2006).
\bibitem{Nicholas} R. J. Nicholas, M. A. Hopkins, D. J. Barnes, M.
A. Brummell, H. Sigg, D. Heitmann, K. Ensslin, J. J. Harris, C. T.
Foxon, and G. Weimann, Phys. Rev. B \textbf{39}, 10955 (1989); J.
Richter, H. Sigg, K. v. Klitzing, and K. Ploog, Phys. Rev. B
\textbf{39}, 6268 (1989).
\bibitem{Merkt} U. Merkt, Phys. Rev. Lett. \textbf{76}, 1134 (1996).
\bibitem{Maksym} P. A. Maksym and T. Chakraborty, Phys. Rev. Lett. \textbf{65},
108 (1990) \bibitem{Wojs} A. Wojs and P. Hawrylak, Phys. Rev. B
\textbf{53}, 10841 (1996).
\bibitem{Savic} I. Savi\'{c} and N. Vukmirovi\'{c}, Phys. Rev.
B \textbf{76}, 245307 (2007).
\bibitem{NTTNguyen1} N. T. T. Nguyen and F. M. Peeters, Phys. Rev. B \textbf{78}, 245311 (2008).
\end{thebibliography}
\end{document}